\documentclass[usenatbib]{mnras}
\usepackage{graphicx,color}
\usepackage{amsmath}	
\usepackage{amssymb}	
\newcommand{\bc}{}

\title[Metallicity dependence of planets]{Metallicity-Dependent Signatures in the {\it Kepler} Planets}
\author[Owen, J. E. \& Murray-Clay R.]{James E. Owen$^1$ and Ruth Murray-Clay$^2$ \\$^1$Astrophysics Group, Imperial College London, Prince Consort Road, London SW7 2AZ, UK; james.owen@imperial.ac.uk\\$^2$Department of Astronomy and Astrophysics, University of California, Santa Cruz, CA 95064, USA; rmc@ucsc.edu}
\begin{document}

\maketitle
\begin{abstract}
{\bc Using data from the California-{\it Kepler}-Survey (CKS) we study trends in planetary properties with host star metallicity for close-in planets. By incorporating knowledge of the properties of the planetary radius gap identified by the CKS survey, we are able to {\bc investigate} the properties of planetary cores and their gaseous envelopes separately.  Our primary findings are that the solid core masses of planets are higher around higher metallicity stars and that these more massive cores were able to accrete larger gas envelopes. 
Furthermore, investigating the recently reported result that planets with radii in the range ($2-6$\,R$_\oplus$) are more common at short periods around higher metallicity stars in detail, we find that the average host star metallicity of H/He atmosphere-hosting planets increases smoothly inside an orbital period of $\sim$20~days. We interpret the location of the metallicity increase within the context of atmospheric photoevaporation: higher metallicity stars are likely to host planets with higher atmospheric metallicity, which increases the cooling in the photoevaporative outflow, lowering the mass-loss rates. Therefore, planets with higher metallicity atmospheres are able to resist photoevaporation at shorter orbital periods. Finally, we find evidence at 2.8$\sigma$ that planets that do not host H/He atmospheres at long periods are more commonly found around lower metallicity stars. Such planets are difficult to explain by photoevaporative stripping of planets which originally accreted H/He atmospheres. Alternatively, this population of planets could be representative of planets that formed in a terrestrial-like fashion, after the gas disc dispersed.   }
\end{abstract}


\section{Introduction}
Both transit \citep[e.g.,][]{KeplerpaperI,KeplerpaperII} and radial velocity \citep[e.g.][]{Howard2010,Weiss2014} surveys have demonstrated that extrasolar planetary systems are structurally diverse.  However, the conditions under which different types of planetary systems form has yet to be established.  Competing planet formation models often predict different dependences on the conditions in the circumstellar disc from which planets form.  Correlations between system structure and stellar properties---which may in turn correlate with those initial conditions---are particularly useful for disentangling different models.  In this paper, we identify new features in the close-in, low-mass planet population that depend on stellar metallicity and discuss their implications for planet formation and evolution.

It has long been known that metal-rich stars are more likely to host close-in giant planets \citep[e.g.,][]{FischerValenti}.  Furthermore, the radial distribution of giant planets around metal-rich hosts shows an enhancement near 3 day periods (``hot Jupiters") that is not seen around low metallicity stars, and in the ``period valley" beyond this 3-day enhancement but interior to $\sim$1AU, high-eccentricity giants are more prevalent around high-metallicity stars \citep{DawsonMurrayClay}.  These features have been interpreted as favouring the core-accretion model for giant planet formation, in which solid material in the protoplanetary disc first produces a solid core with a mass substantially exceeding that of the Earth, and the core then accretes a massive gas envelope 
\citep[e.g.,][]{Pollack1996,Rafikov2006}.  More metal-rich stars presumably hosted discs with more solid material, making the accumulation of large solid cores easier.  Beyond $\sim$1AU, the occurrence rate of giant planets increases, suggesting that most period valley giants may be dynamical transplants from originally more distant orbits.   \citet{DawsonMurrayClay} suggest that the metallicity-dependence of {\bc the abundance of} period valley Jupiters indicates that dynamical interactions between giant planets---formed more readily around high-metallicity stars---may have {\bc put in place} the majority of the close-in giant population.  

A straightforward consequence of this interpretation of the metallicity-dependence of giant planet frequency is the expectation that the sizes and masses of smaller planets {\bc could} also depend on the metallicities of their host stars.  In other words, if a high-metallicity star is more likely to host a disc with enough solid material to produce a solid core exceeding the threshold required to accrete a massive gas envelope (e.g. $M_{\rm core}\sim 10M_\oplus$), {\bc one might hypothesise} the final masses of cores that do not exceed that limit---instead remaining as smaller planets---to be smallest for the lowest metallicity stars.  Alternatively, this correlation could be altered by redistribution of solid material due to radial drift of small solids through the protoplanetary gas disc \citep{Weidenschilling1977}.  When present, gas giants open gaps in their host discs, producing pressure maxima that can trap inwardly-drifting solids \citep{Rice2006}, potentially reducing the supply of solid material to the inner disc \citep[e.g.,][]{Morbidelli2016}.  If close-in {\bc planets} reflect formation from solid material that drifted into the inner disc because no giant planets were present to halt drift, {\bc these planets} should be more common and more massive around lower-metallicity stars that could not produce giant planets.  

Observationally, however, evidence of metallicity-dependence of either sign has proven elusive. There is some evidence from the LAMOST survey that very close-in planets (periods $<$ 10 days) are more common around higher metallicity stars \citep{Mulders2016} including the fact hot Neptunes are more common around higher metallicity stars \citep{Dong2017}. However, small, ultra-short period planets (periods $<$ 1 day)  do not show evidence of a host star metallicity preference at the same level as hot Jupiters \citep{Winn2017}. \citet{Petigura2018} has recently demonstrated that there is a greater diversity of planets around higher metallicity stars, and intriguingly that the short-period break in the period dependence of the small  planet ($\lesssim 1.7$ Earth Radii) occurrence rate moves to shorter periods around higher metallicity stars. \citet{Petigura2018} also presented the absolute occurrence rates for different populations of planets with orbital periods shorter than 100~days in several metallicity bins, showing in the majority of cases planet occurrence increased with stellar metallicty. However, the population of close-in planets {\bc with radii $\lesssim 6$~R$_\oplus$} discovered by NASA's {\it Kepler} mission have measured radii that do not clearly depend on metallicity and masses are not yet available for a large enough population of these small planets to conduct an analysis on. 

Recent work by \citet{Fulton2017} provide a clue to the reason for this lack of a direct correlation between stellar metallicity and planet radius.  Using data from the California-Kepler Survey (CKS -- \citealt{CKSI, CKSII}), which provided precise stellar radii for a sample of stars for which planets have been discovered in transit by NASA's \textit{\it Kepler} mission, they show that the radius distribution for these planets is bimodal, confirming predictions that the atmospheres of these planets are strongly shaped by XUV-driven photoevaporation \citep{OwenWu13,Lopez2013}.    The dip arises because small complements of atmospheric hydrogen substantially increase a planet's radius---the low atmospheric masses required to produce radii within the observed dip would be quickly photoevaporated away; whereas larger atmospheres are harder to evaporate and more stable. Specifically, the photoevaporative mass-loss time for volatile atmospheres is longest at the point when they double the core's radius.  \citet{OwenWu17} and \citet{Jin2017} demonstrate that the planetary radius distribution observed by \citet{Fulton2017} as a function of orbital period is well-matched by the ``evaporation-valley'' model.  In other words, the radius distribution at these short periods is shaped more by photoevaporation than by planet formation processes. This conclusion has recently been tightened by \citet{vaneylen2017} using a set of planets for which the stellar parameters could be determined to high precision using asteroseismology. \citet{vaneylen2017}, not only confirmed the presence of a deep gap in the radius distribution to high significance at the same location as \citet{Fulton2017}, but also showed that the planetary radius at which the gap appeared decreases with increasing orbital period, as predicted by photoevaporation models \citep{OwenWu13,Lopez2013,OwenWu17}.

With this complicating physics in mind, in this paper we search the CKS sample for evidence of metallicity-dependence in the populations of very close-in stripped cores and of wider-separation planets that are not expected to experience substantial atmospheric loss due to photoevaporation.  Because the process of photoevaporation is itself likely metallicity-dependent, we further examine the population of close-in planets for evidence of metallicity-dependent atmospheric loss.  Because the populations of giant planets have been better sampled in earlier work, we focus on small planets with radii $<6$R$_\oplus$, where R$_\oplus$ is the radius of the Earth.  We find evidence that stellar metallicity indeed correlates with planetary system properties, providing insights into both the formation mechanism and how they evolve. We note that since the CKS survey only targeted and determined the metallicity of stars that host planets we cannot make statements about how the overall {\bc absolute} occurrence rate varies with metallicity, rather we focus of relative differences.


\section{Overview}\label{sec:overview}

Throughout this analysis, we use data from the CKS survey, as reported in \citet{CKSI,CKSII,Fulton2017} and retrieved online\footnote{https://github.com/California-Planet-Search/cks-website}. We follow exactly the same sample selection criteria as \citet{Fulton2017}: planet candidates previously identified as false positives were removed; we restrict ourselves to bright stars which have Kepler magnitudes $<14.2$, effective temperatures in the range 4700-6500~K and are not giants\footnote{See Eq~1. of \citet{Fulton2017} for this empirical filter.}; finally we choose planets that have impact parameters less than 0.7 and periods shorter than 100 days. This restricted sample results in 900 well characterised planets. By choosing exactly the same planet sample as \citet{Fulton2017} we can follow their procedure for correcting the planet candidates for incompleteness using the published detection probabilities for this sample. Our one dimensional distributions (e.g. histograms) have the incompleteness corrected in the identical way to that described in \citet{Fulton2017}.  Our two-dimensional planet occurrence distributions are estimated using weighted kernel density estimation \citep[wKDE, e.g.][]{MortonSwift} with a bivariate Gaussian kernel. {\bc The width of this Kernel is determined from the uncertainty in the measurement, but we adopt a minimum value to produce smooth disturbitions. Unless otherwise stated we choose a minimum bandwidth of 10\% of the measured value}. In all cases below we use stellar and planetary properties derived from the combination of CKS spectral measurements \citep{CKSI} and isochrone fitting \citep{CKSII}\footnote{The same features presented here are also present when using the metallicities determined purely from the CKS spectra.}. {\bc Finally, even though we preform an incompleteness correction, at small planet radii the finite sample size leads to sampling errors (e.g. a single planet can be detected in a region of low completeness leading to a very large weight). The completeness curve is period dependent: it is easier to detect smaller planets closer to the star. \citet{Fulton2017} provide the completeness curves of the CKS sample. We find a sample completeness of $\sim 25\%$ provides a balance between reducing sampling errors and maintaining a large sample size. Later in this paper we will split the sample into two period ranges $<2.5$~days and $>25$~days which gives us minimum planetary radii of 0.6~R$_\oplus$ and 1.0~R$_\oplus$ respectively. We have checked small departures from these values do not effect our results.} As will become clear the radius gap reported by the CKS team \citep{Fulton2017} is of critical importance to interpreting the metallicity trends. We take the gap to occur at 1.8 R$_\oplus$ at all orbital periods, and refer to planets with radii greater than this as ``large'' planets and those smaller than this as ``small'' planets.     

\begin{figure}
\includegraphics[width=\columnwidth]{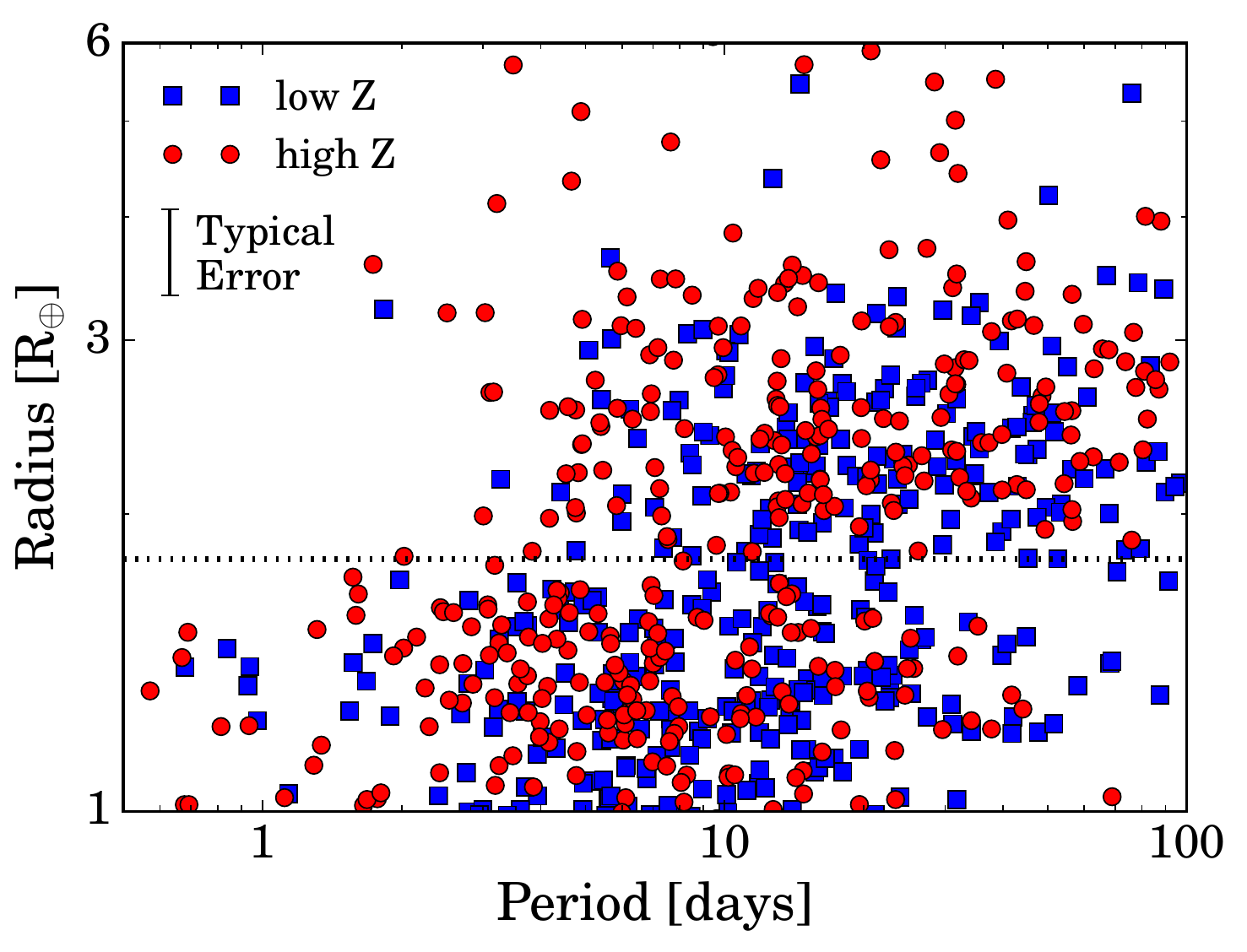}
\caption{The radii and periods of well characterised Kepler planets. The sample is split into two populations based on the host stars metallicity. Blue squares show planets around stars lower than the sample's median metallicity of $[M/H]=0.046$, whereas red circles show those planets whose host star's metallicity is above the median value. The position of the radius gap is shown as the dotted line.}\label{fig:overview}
\end{figure}

Figure~\ref{fig:overview} shows the radius and period of the CKS planets where we have split the sample in terms of high and low metallicity about the median value of $[M/H]=0.046$ {\bc (to produce to equally sized sub-samples)} with the radius gap indicated by the dashed line. Even without correcting for the incompleteness several features are immediately obviously in Figure~\ref{fig:overview}. Firstly, it is clear large planets reside at shorter periods around higher metallicity stars, such a result has been identified before  using LAMOST spectra \citep{Dong2017} and by the CKS team themselves \citep{Petigura2018}. Secondly, large planets with radii $\gtrsim 3~$R$_\oplus$ are predominately around high metallicity stars {\bc as also noted by \citep{Dong2017}}. Finally, small planets at long periods appear more common around low metallicity stars.


\begin{figure*}
\centering
\includegraphics[width=\textwidth]{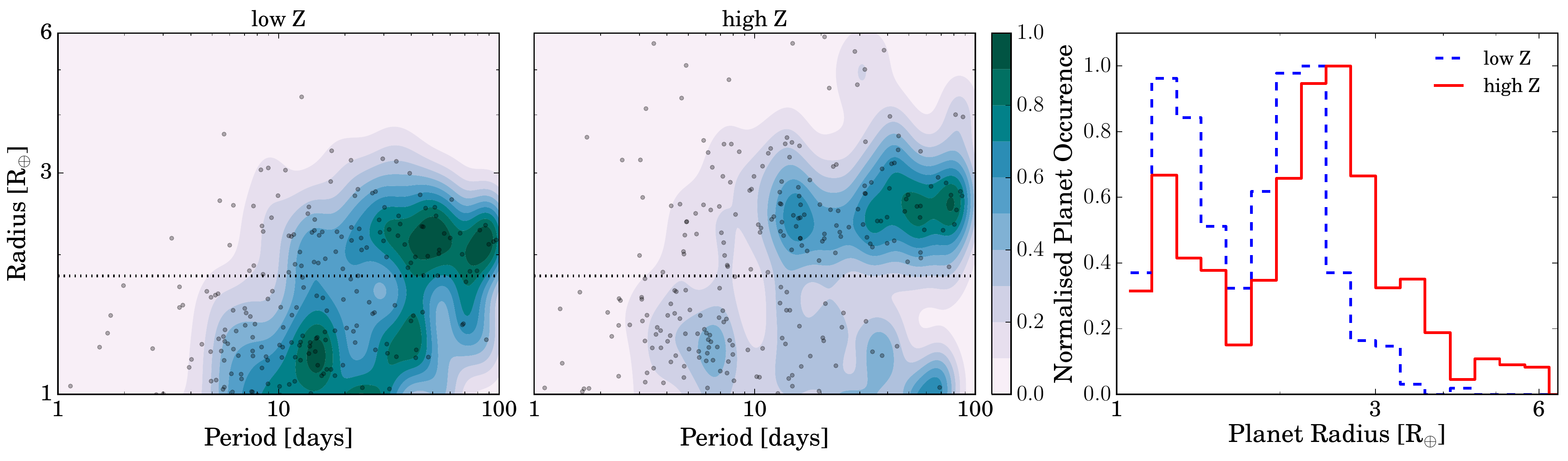}
\caption{The radius-period distribution and radius distribution for the bottom third ($[M/H]\le-0.023$) of the metallicity distribution and the top third ($[M/H]\ge0.0993$) of the metallicity distribution. To provide greater smoothing to the radius-period distribution, the minimum bandwidth of the wKDE is set to 15\% of the points value in the two radius-period distributions. {\bc The histograms in the right-hand panel are normalised such that they both peak at unity.}} \label{fig:rad_period_met_bins}
\end{figure*}

Once we correct for the observational incompleteness we can inspect these trends in detail. Figure~\ref{fig:rad_period_met_bins} shows the radius-period distributions of planets around low (bottom third) and high (top third) metallicity stars, as well as the radius distribution. The radius distribution shows us that not only are planets with radii $\gtrsim 3\,$R$_\oplus$ more common around higher metallicity stars, but that the radius distribution of {\it all} large planets is shifted to larger radii around higher metallicity stars, i.e. large planets are larger around higher metallicity stars. The radius-period distributions clearly show that larger planets can reside closer to their host stars around higher metallicity stars. Furthermore, we find no evidence that either the gap in the radius distribution or the valley in the radius-period distribution is metallicity dependent. The change in gap position in the radius distribution is particularly stringent and we find it does not change by  $\gtrsim 15$\% in radius over a wide range of host star metallicity. 

\begin{figure}
\includegraphics[width=\columnwidth]{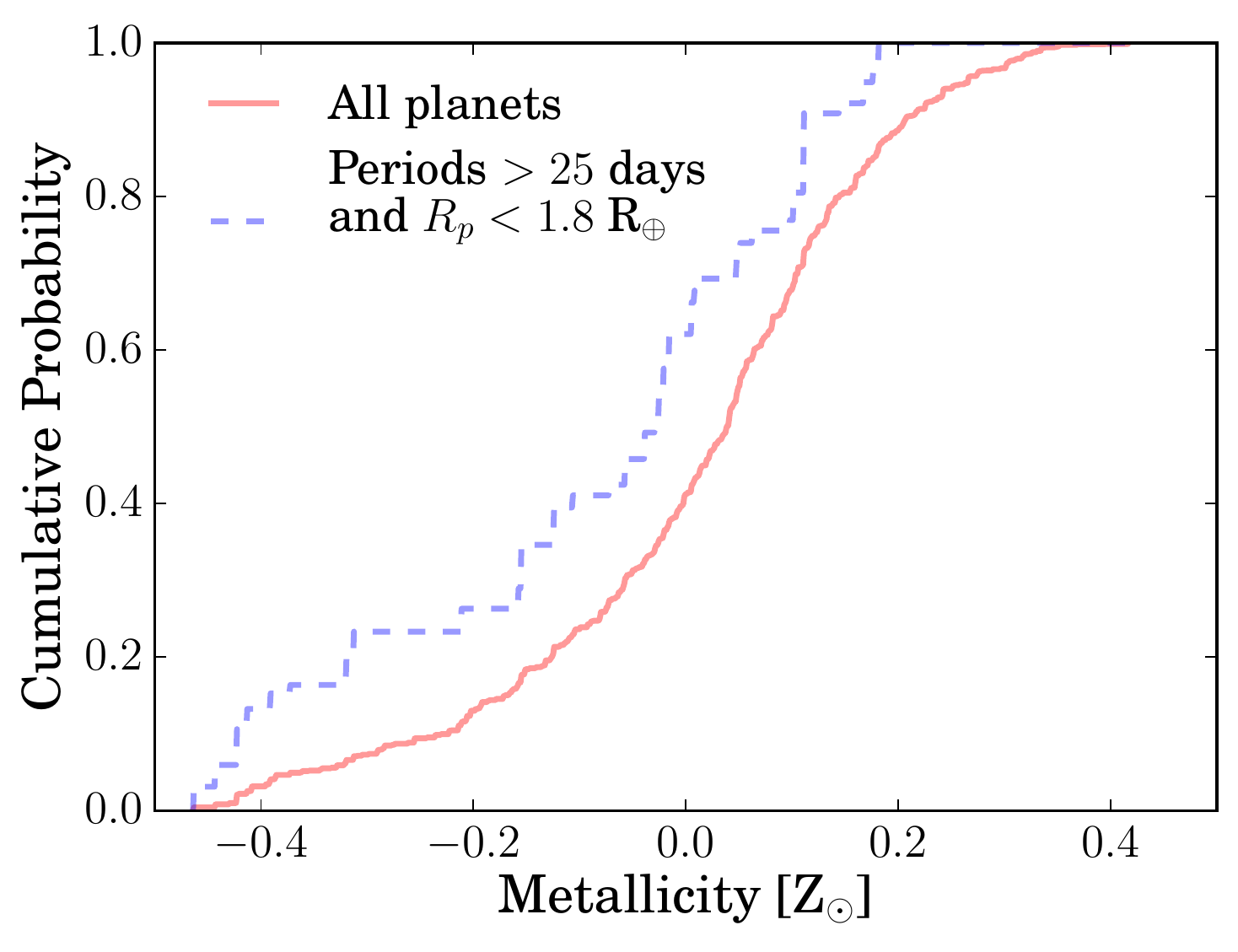}
\caption{Metallicity distribution for the full CKS sample of planets with radii $> 1\,$R$_\oplus$ (solid) and the sub-sample for which the orbital period is greater than 25 days and the planetary radius is less than 1.8$R_\oplus$ (dashed).  Small planets at large periods are preferentially found around low-metallicity stars.}
\label{fig:long_period_met}
\end{figure}

Finally, even at large periods, the ratio of small to large planets appears larger around low-metallicity stars (Figure~\ref{fig:rad_period_met_bins}).  We expand upon this observation in   Figure~\ref{fig:long_period_met}, which shows the metallicity distribution of stars hosting small (1-1.8\,R$_\oplus$) planets with periods $>$ 25 days compared to the metallicity distribution of stellar hosts for all planets with radii $>1\,$R$_\oplus$ (i.e., 1-6R$_\oplus$). Small planets with long periods appear to be more common around lower metallicity stars. We use a {\bc $k$-sample Anderson-Darling test, using the right-side empirical distribution function \citep{Scholz1987} to compare our distributions.} Comparing the host star metallicity distribution for small, long-period planets to that for all planets reveals a test statistic of 5.94. This test statistic tell us that the probability that these two distributions are drawn from the same underlying one is $5.4\times10^{-3}$ and thus we can reject this hypothesis with $\sim2.8\sigma$ confidence\footnote{As our data is weighted, {\bc due to incompleteness corrections}, the standard method available in common software packages for turning a k-sample Anderson-Darling test statistic into a significance value is not appropriate for finite samples. {\bc Therefore, we use Monte-Carlo re-samplings of the data to repeat the test 50,000 times to measure the significance.}}. This hypothesis test indicates that there is good evidence that small, long-period planets are indeed more common around lower metallicity host stars. We note that small planets at long periods are the most difficult planets to observe in this sample, making a careful evaluation of selection effects important, as discussed in Section \ref{sec:caveats}.



\section{Signatures of Planet Formation}\label{sec:formation_sig}
While the evaporation-valley's theoretical explanation of the observed planet radius gap is the best theory we have to explain its origin, other explanations for this feature may be possible \citep[e.g.][]{Ginzburg2018}.  Regardless, there is no doubt that close-in planets can lose mass due to atmospheric escape. Therefore, to investigate signatures of planet formation we must be careful to negate any possible effects that photoevaporation may have had over the observed planets' billion year histories. This especially imperative since photoevaporation is expected to be metallicity-dependent \citep{OwenJackson}. Atmospheric evaporation depends on the XUV ``exposure'' of the planet \citep[e.g.][]{Lecavelier2007}, i.e. the total XUV energy the planet receives over its lifetime.  A star's XUV luminosity varies with stellar mass and age. Empirical comparisons indicate that orbital period is a better directly observable proxy compared to the current bolometric flux or X-ray flux for a planet's history of XUV illumination \citep{OwenWu17}. {\bc This is because at fixed period the drop in bolometric flux with stellar mass is {\it approximately} counterbalanced by the increase in $L_{\rm XUV}/L_{\rm bol}$}. 

By cutting the sample into those with long orbital periods and separately those with short orbital periods we can investigate the properties of planets that have either experienced little mass-loss during their lives or those which have experienced complete evaporation.  For planets with periods $>25$ days theoretical models predict evaporation is minimal \citep[e.g.][]{Lopez2013,OwenWu13,Erkaev2016}. {\bc Specifically, the model of \citet{OwenWu17} which was built for the CKS planet sample found that at periods of $>25$ days planets with masses $\gtrsim 2$~M$_\oplus$ were stable against photoevaporation for a range of mass-loss prescriptions and core compositions}. Therefore, for planets which host H/He envelopes, looking at large periods allows us to {\bc study} the {\bc properties of atmosphere hosting planets at formation}. {\bc This inference of course assumes that no other processes strongly effects a planet's evolution after formation at long-periods.} Conversely, at very short periods evolutionary models predict that evaporation is so efficient that low-mass planets are unable to retain any primordial H/He envelopes. Looking at short periods allows us to investigate the properties of the solid cores alone, independent of how much H/He they accreted initially. Therefore, we also investigate the population of planets with periods shorter than 2.5 days, where evolutionary calculations and the data itself (there are almost no planets above the gap) suggest that no planets have H/He envelopes. 

\begin{figure}
\includegraphics[width=\columnwidth]{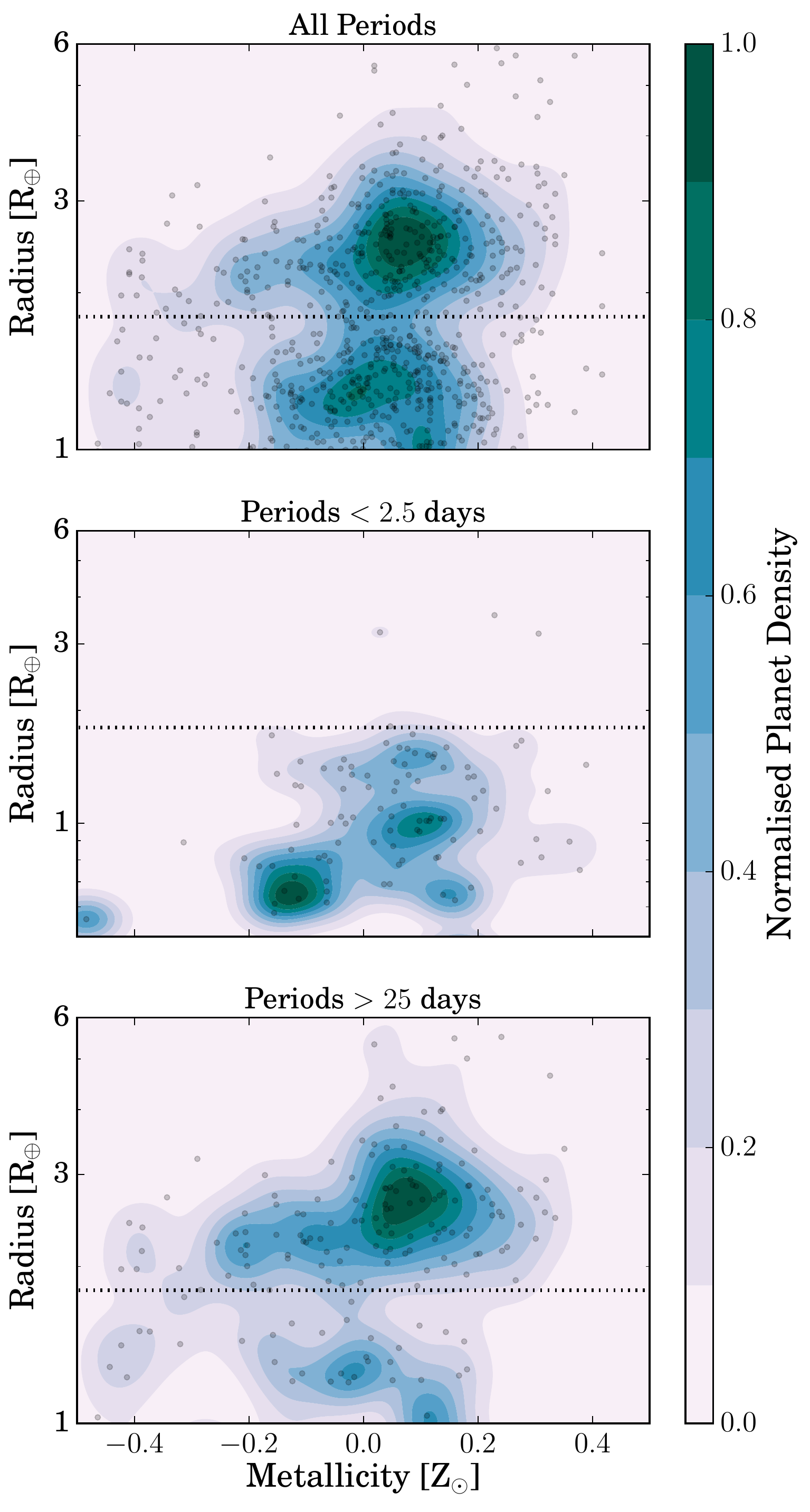}
\caption{The radius-metallicity distribution of the CKS planets. The top panel shows all planets, the middle panel shows those planets with periods shorter than 2.5 days and the bottom panel shows planets with periods longer than 25 days. The dotted line shows the location of the radius gap. The $y$-axis extends to smaller planetary radii for the middle-panel as the sample completeness is higher for smaller planets close to their stars -- see Section~\ref{sec:overview}.}\label{fig:period_cut_panels}
\end{figure}

\begin{figure}
\includegraphics[width=\columnwidth]{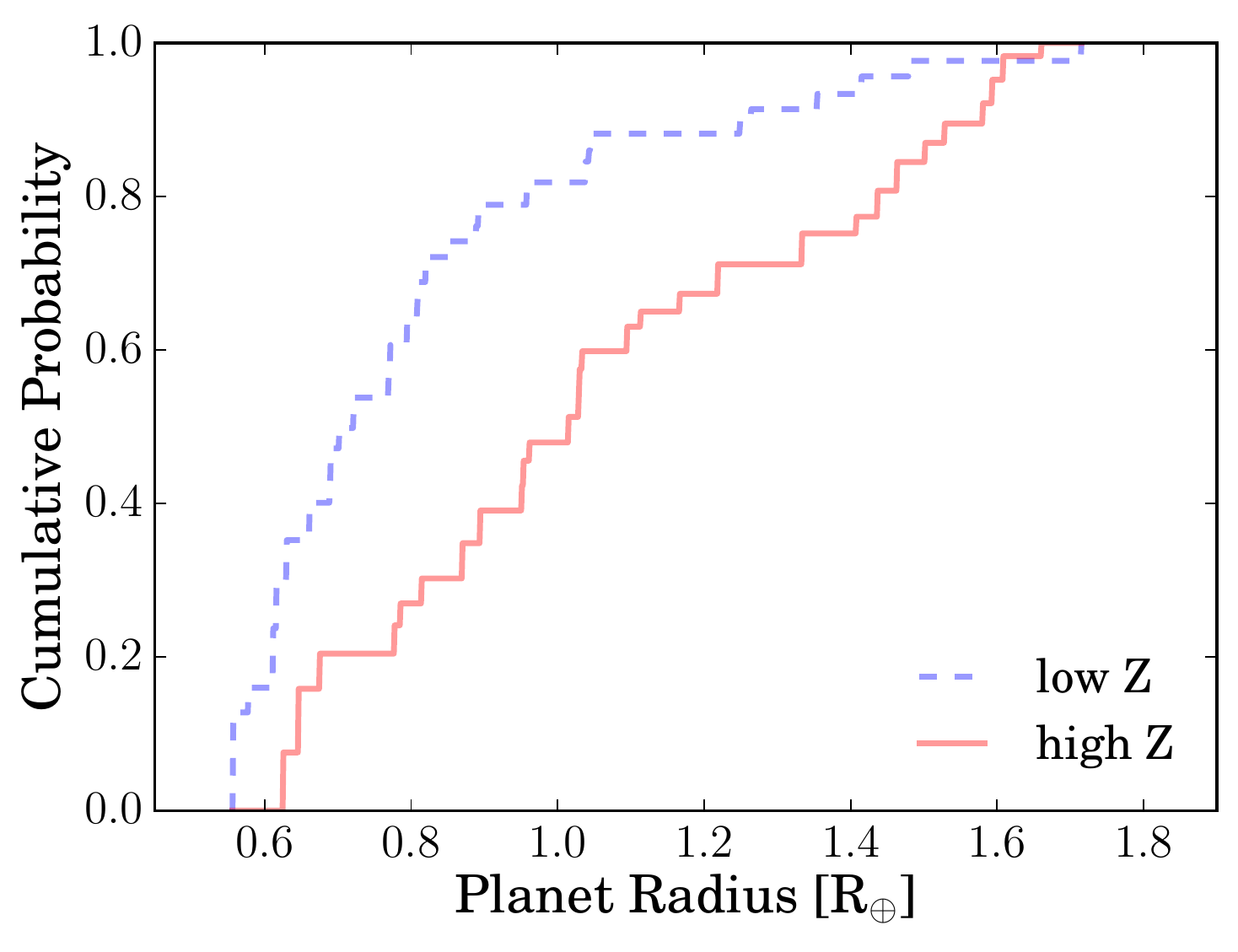}
\caption{The radius distribution of small planets with periods shorter than 2.5 days. The sample has been split into two about the median metallicity with planets around low metallicity stars shown as the blue dashed line and planets hosted by high metallicity stars shown as the red solid line. It is clear lower metallicity stars host smaller, short period planets.}\label{fig:small_short_planets}
\end{figure}

Figure~\ref{fig:period_cut_panels} shows the planet radius-metallicity distribution for all planets and those with periods shorter than 2.5 days and those longer than 25 days. It is obvious that the trend identified earlier: that large planets are larger around higher metallicity stars is true at long periods. The fact that this trend exists for planets at long periods indicates that it is an imprint of formation rather than the outcome of photoevaporation. We also see that for short-period small planets that they are larger around higher metallicity stars. This trend is investigated further in Figure~\ref{fig:small_short_planets}, where we plot the radius distribution for short period planets in two sub-samples: those hosted by low metallicity stars and those hosted by high metallicity stars. It is clear that low metallicity stars host small planets with smaller radii in the CKS sample. We perform a {\bc k-sample} Anderson-Darling test to test the significance of this result finding a test statistic of 5.58, indicating that the probability\footnote{As before this is determined using a Monte-Carlo re-sampling.} that these two samples are drawn from the same underlying one is $5.9\times10^{-3}$, allowing us to reject this hypothesis with $\sim2.8\sigma$ confidence. Interpreting this result in terms of the properties of the solid cores of low-mass planets (because their gas has been stripped), we see that the solid cores that form around lower metallicity stars are typically smaller and therefore less massive, {\bc assuming they have similar compositions}. 

Alternatively, at large periods we find no evidence that the radius distribution of small planets is metallicity dependent. This result can be seen in the bottom panel of Figure~\ref{fig:period_cut_panels}; inspection of the radius distributions for small long-period planets in different metallicity bins yields no evidence for a metallicity dependence at any believable significance, although we note the small sample size in this case. Therefore, small planets at long periods, while more frequent around low metallicity stars, appear to have a radius distribution that is not strongly sensitive to metallicity. We suggest possible explanations for this insensitivity in Section \ref{sec:discuss}.

Connecting the results for large planets at long periods with small planets at short periods we can conclude that forming planets around higher metallicity stars build larger (higher mass) cores, due to the presumably higher solid content available. This is evidenced by the radius distribution of planets at short periods where evaporation has exposed the stripped cores, {\bc showing larger (and likely more massive) cores around higher metallicity stars}. These more massive cores are then able to accrete larger gaseous envelopes due to their deeper gravitational potential wells as indicated by the properties of large planets at long periods where photoevaporation has left their gaseous envelopes untouched. These results are in agreement with our qualitative understanding of how planets form and acquire a gaseous envelope through a core-accretion-like mechanism. 

While one requires an understanding that evaporation happens and can affect a planet's evolution, we emphasise the above results are actually independent of whether the gap's origin is truly the evaporation-valley scenario or not.

\section{Signatures of Photoevaporation}\label{sec-smallP}

Metallicity can have an important effect on a exoplanet's hydrodynamic photoevaporative outflow. Atomic cooling lines from elements such as Oxygen and Carbon are important for XUV heated gas with temperatures of a few thousand to ten-thousand Kelvin. Therefore, increased metallicity tends to result in cooler gas temperatures at a fixed density. This consequence means the evaporative flow has more difficulty escaping the planet's potential, resulting in lower-mass loss rates.  

This is true for UV heated gas where the heating rate is dominated by photo-ionization of Hydrogen and is therefore roughly independent of the metal content, but the cooling rate increases with increasing metallicity. This is also true in the case of X-ray heated gas; while the heating originates from K-shell photo-electrons produced from metals (a process for which the rate is $\propto n_Z$), many of the metal cooling lines are collisionally excited and therefore have a steeper dependence on the metal content than the heating rate. There are very few real calculations of the effect of the metal content of a planet's upper atmosphere on evaporation. However, there has been some work on the photoevaporation of accretion discs \citep[e.g.][]{EC10}. Borrowing a global scaling from work on the X-ray heating of protoplanetary disc atmospheres \citep{EC10}, \citet{OwenJackson} suggested that the photoevaporative mass-loss rate from exoplanets scaled with metallicity $Z$ as $Z^{-0.77}$. It is important to emphasize that the exact scaling between metallicity and planetary evaporation rates is yet to be explored properly in the context of exoplanets and the global scaling found by \citet{EC10} in the accretion disc context is now known, unsurprisingly, not to hold locally \citep{Ercolano2017}. {\bc Furthermore, if there is a metallicity gradient in the planet's atmosphere, then this will also affect how the planet evolves, perhaps with the atmospheric metallicity increasing with time.} However, the basic physics underpinning the fact that photoevaporative mass-loss rates should fall with increasing metallicity is true, provided that the metal abundance is not so low that Lyman-alpha and Hydrogen recombination are the dominant cooling mechanisms. There is a further obvious point but it is worth making explicit: one measures the metallicity of the host star, rather than the metallicity of the exoplanet's envelope. Since both the star and planet accrete gas from the same circumstellar disc, there is every expectation that they should correlate, but processes such as planetesimal ablation and pebble accretion mean this does not have to be a uniform or linear mapping. 

The fact that large planets reside closer to their stars around higher metallicity stars (Section~\ref{sec:overview}; Figure~\ref{fig:overview}) is good evidence that the complete stripping of primordial atmospheres is harder from planets with higher metallicity envelopes (that presumably accreted a higher metal content when forming around a higher metallicity star). {\bc Such a trend is difficult to produce from atmospheric accretion of nebula gas as  forming planets with higher metallicity envelopes cannot cool efficiently and subsequently accrete less gas (for fixed core mass) \citep[e.g.][]{Rafikov2006,Lee2015}. However, this slower cooling can  be overcome  if the metallicity enrichment is so large that it increases the mean-molecular-weight of the gas sufficiently to decrease the scale height of the atmosphere allowing more massive atmospheres. The cross-over atmospheric metallicity is around $Z\sim0.2$ \citep{Lee2015}. Such envelopes metallicities can be reached in the case of accretion of ice-rich planetesimals which enrich the envelope metallicity. For example, \citet{Venturini2016} showed that cores massive enough to completely disrupt icy planetesimals (where the water remains mixed within the envelope) could reach envelope metallicities above $0.2$. The calculations of \citet{Venturini2016} are focused on core accretion outside the snow-line, it is unclear if this is applicable to the close-in planets considered here, or what happens in the case of accretion of refractory rich solids.}  Another possible explanation of the fact large planets reside closer to their stars around higher metallicity stars  is simply higher core masses being produced around higher stellar metallicity stars. 

For example, for a passively cooling atmosphere the envelope mass fraction, $X$ scales with metallicity and core-mass as \citep[e.g.][]{Lee2015}, {\bc for $Z\lesssim 0.2$}:
\begin{equation}
X\propto Z_{\rm env}^{-b/(2+\alpha)}M_c^{2/3(\frac{1+\alpha}{\gamma-1}-1)/(2+\alpha)}
\end{equation}
where $b$ and $\alpha$ are power-law opacity scalings with metallicity and pressure respectively and $\gamma$ is the ratio of specific heats. Adopting $b=1$ and $\alpha=0.5$ as is appropriate for the opacity arising from H$^{-}$ suggested to dominate at the radiative-convective boundary (similar to the $\alpha=0.68$ found by \citealt{Rogers2010} when performing global power-law fits to the \citealt{Freedman2008} opacity tables), and $\gamma=1.2$ \citep[see][]{Lee2015}, we find $X\propto Z_{\rm env}^{-0.4} M_c^{1.73}$. Furthermore, if we take the core's mass-radius relationship to follow $M_c\propto R_c^4$, appropriate for rocky bodies \citep[e.g.][]{Valencia2006,Fortney2007,Lopez2014}, and writing $M_c\propto Z_{\rm env}^c$ {\bc (where the parameter, $c$, encodes the variation of core mass with envelope metallicity)} we can derive a constraint on how strongly the core mass needs to increase with metallcity to overcome the slower cooling of higher metallicity envelopes.  Thus, for the envelope mass-fraction to increase with the envelope metallicity\footnote{Obviously this limiting value for $c$ is sensitive to the choice of mircophysical parameters, and the general result requires $c>b/\{1/4[3(\alpha+1)/(\gamma-1)-\alpha]-1\}$} one requires $c\gtrsim 0.2$.

After the planetary envelope has finished forming and it begins its evolution, a higher metallicity envelope will cool slower and therefore make it larger at a fixed age and hence easier to evaporate.  A simple analysis for how the mass-loss timescale ($t_{\dot{m}}\equiv M_{\rm env}/\dot{m}_w$) of a passively cooling adiabatic envelope, with a fixed core mass (see Equation 16 of \citealt{OwenWu17}), scales with {\it envelope} metallicity ($Z_{\rm env}$) indicates that:
\begin{equation}
t_{\dot{m}}\propto Z_{\rm env}^{a-b/(2+\alpha)}
\end{equation}
where $a$ is the index for how the mass-loss rate scales with metallicity ($\dot{M}\propto Z^{-a}$). For $a=0.77$ and again adopting $b=1$, $\alpha=0.5$,  we find that $t_{\dot{m}}\propto Z_{\rm env}^{0.37}$ for fixed core-mass. For fixed {\bc envelope} metallicity, more massive cores accrete larger envelopes \citep[e.g.][]{Rafikov2006}, and since higher metallicity stars result in more massive cores (Figure~\ref{fig:small_short_planets}), it is clear that the mass-loss timescale for planet's with more metal rich envelopes is likely to be longer. This inference is consistent with what we see in the planet population, assuming, of course, higher metallicity stars yield higher metallcity planetary envelopes.  
Measuring the mass of close-in, large planets {\bc may} help untangle the relative importance of core-mass and atmospheric metallicity in driving this trend. 




\begin{figure}
\includegraphics[width=\columnwidth]{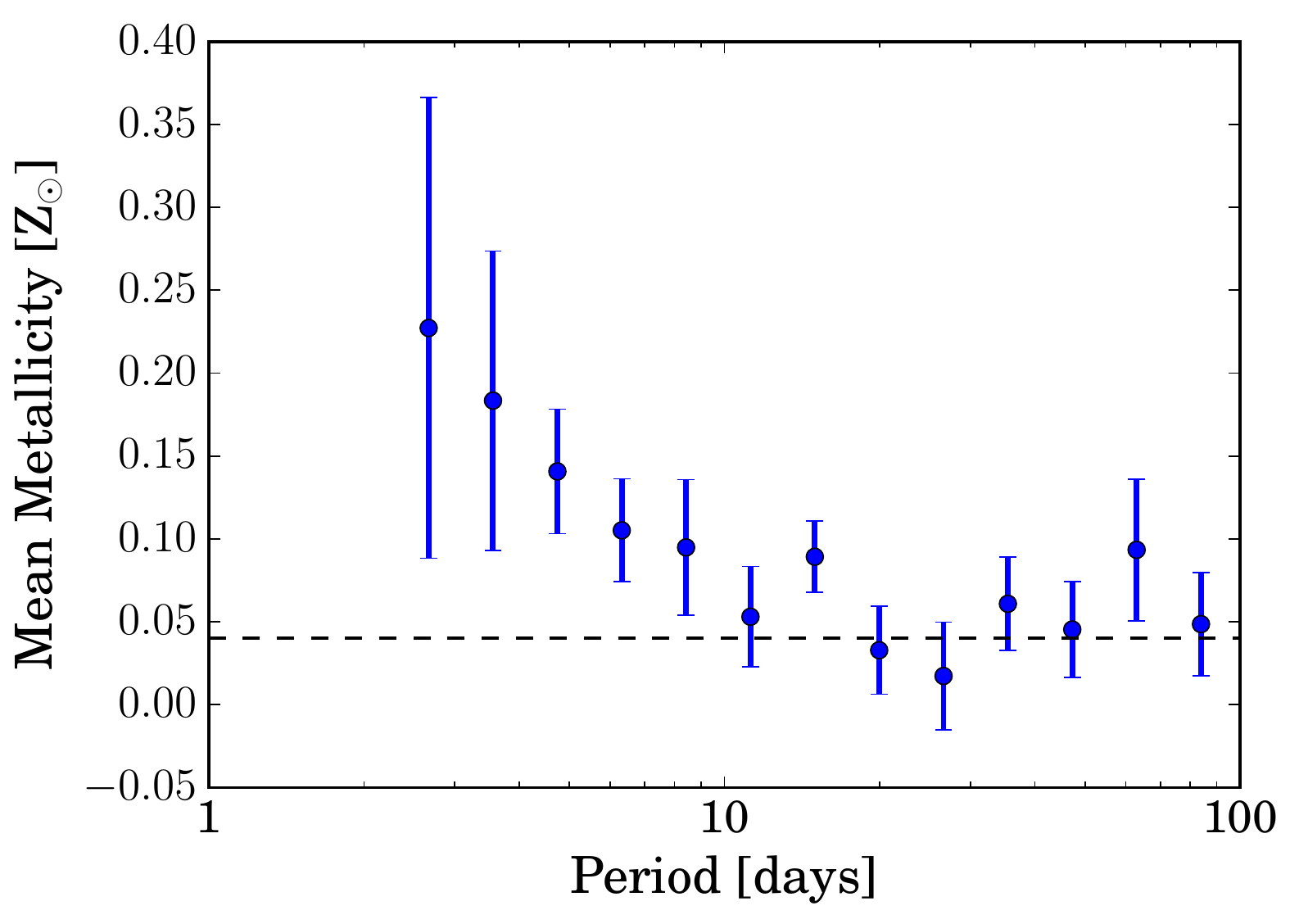}
\caption{The mean metallicity of stars which host large planets {\bc (1.8 $< R_p <$ 6~R$_\oplus$)} as a function of period. The dashed line shows the weighted mean metallicity of the CKS sample. Since it is not possible to uniquely define standard errors in the mean for weighted data, our error bars are calculated using formula provided by \citet{Cochran1977}, as recommended by \citet{Gatz1995}.}\label{fig:Av_met_period}
\end{figure}


We can look at this effect more quantitatively by investigating the average host star metallicity of large planets as a function of orbital period. This is shown in Figure~\ref{fig:Av_met_period}. We see that at large periods the average host star metallicity is consistent with the bulk of the population in the CKS sample.  This is as expected as large planets are more common at long periods, but it is also independent of period from 20-100 days. Inside, 20 days the average host star metallicity of the large planets begins to smoothly increase. The period at which this increase begins to happen is exactly where theoretical models predict evaporation starts to have evolutionary consequences. 

\citet{Dong2017} suggested that the similarity that close-in large planets and hot jupiters are present around higher metallicity stars points to a common origin, like high-eccentricity migration. The fact the average metallicity around large planets starts to increase around 20 days is further evidence that this is unlikely, as even for a planet with a tidal quality factor of 100, the circulations time for a 10 M$_\oplus$, 3 R$_\oplus$ is $>$ 10 billion years at 20 days. Given that large planets probably have large H/He envelopes, a tidal quality factor of $\sim$100 is incredibly optimistic.  

Therefore, this result should be seen as empirical evidence for the role atmospheric metallicity can play in atmospheric escape and should stimulate more theoretical work in this area to explicitly model the role of heavy elements in cooling photoevaporative flows. 


\section{Potential Caveat: Stellar mass vs. metallicity}\label{sec:caveats}
\begin{figure}
\includegraphics[width=\columnwidth]{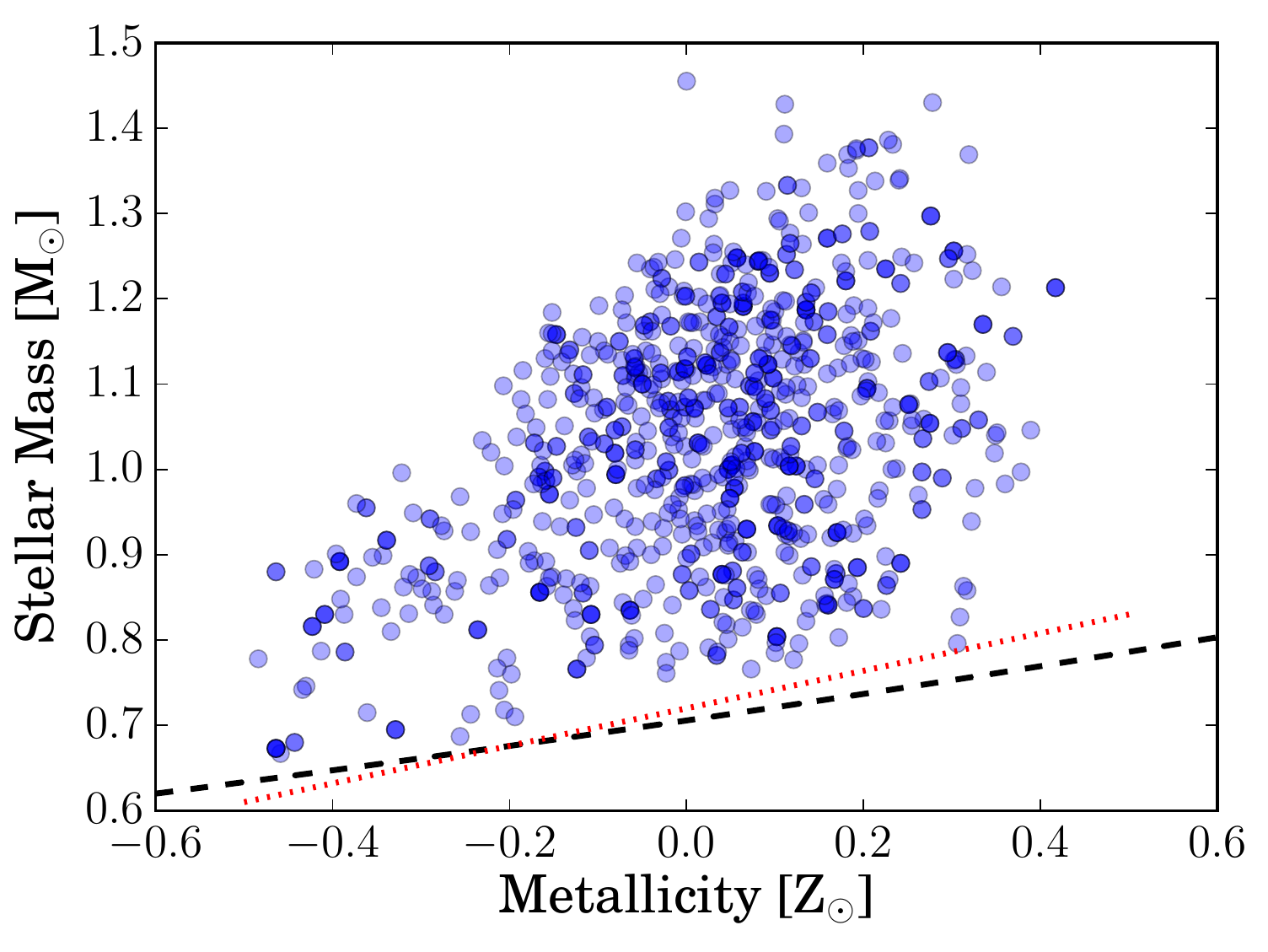}
\caption{The points show the masses and metallicities of all 900 planet host stars in the CKS sample of \citet{Fulton2017}. The dashed line shows the theoretically expected cut-off below which the flux-limited sample will not include stars due to the fact lower metallicity stars are more luminous. The dotted line shows the expected cut-off as stars with effective temperatures $<4700$~K are excluded from the sample.}\label{fig:star_metal}
\end{figure}

There is an important caveat to all the trends determined above. The stars in the reduced CKS sample were obtained through a flux-limited sample {\bc for stars with effective temperatures $>4700$~K} and as such are prone to biases. { \bc These biases result in two cut-offs that remove high-metallicity, low-mass stars from the sample. Firstly, and most importantly is the fact the spectroscopic fitting tools used by the CKS team \citep{CKSI} exclude stars with effective temperatures $<4700$~K. Secondly there is a Malmquist bias arising from the fact lower metallicity stars are more luminous. In Figure~\ref{fig:star_metal} we show the metallicity and stellar mass of all 900 stars in the reduced CKS sample of \cite{Fulton2017}. The dashed line shows the expected Malmquist bias based on a Kepler-magnitude limited sample and the dotted line shows lower boundary arising from the effective temperature boundary. These limits were derived using scaled-solar {\sc mesa} stellar evolution models taken from the online MIST tracks for stars with an age of 3~Gyr \citep{MESAI,MESAII,MESAIII,MIST0,MISTI}.} 
It is not clear to us what is driving the upper envelope with possible suggestions being galactic evolution: higher-mass stars have shorter lives and are therefore more likely to have formed in a metal rich environment if we can see them today, producing an age-metallicity relation \citep[e.g.][]{Garnett2000}; galactic structure: looking towards higher metallicity environments (e.g. spiral arms), more distant and more massive stars would be in the higher metallicity environment if the metallicity distance increases with distance from the Sun in the direction of the Kepler field; or a combination of a $v\sin i$ and effective temperature cut-off.

Origins of the correlation between stellar mass and metallicity aside, the more important question is whether the trends discovered above truly driven by metallicity or just a proxy for a stellar mass effect. The fact that stellar mass and metallicity correlate positively means that the same theoretical expectation of more solids gives bigger planets applies equally to more massive or more metal rich stars. This is because higher mass stars have more dust mass in their protoplanetary discs \citep[e.g.][]{Ansdell2016,Pascucci2016}. Therefore all the conclusions and implications on the planet formation process identified in this work are safe if one is really interested in how planet formation varies with the protoplanetary disc's total solid content. 

Nevertheless, to investigate the trends identified in Section~3 we take a narrow stellar mass range from 0.85 to 1.15~M$_\odot$ {\bc and exclude those planets with host star metallicities $<-0.2$}.  In this sub-sample there is no strong mass-metallicity trend and reproduce Figure~\ref{fig:period_cut_panels} in this mass range.  
\begin{figure}
\centering
\includegraphics[width=\columnwidth]{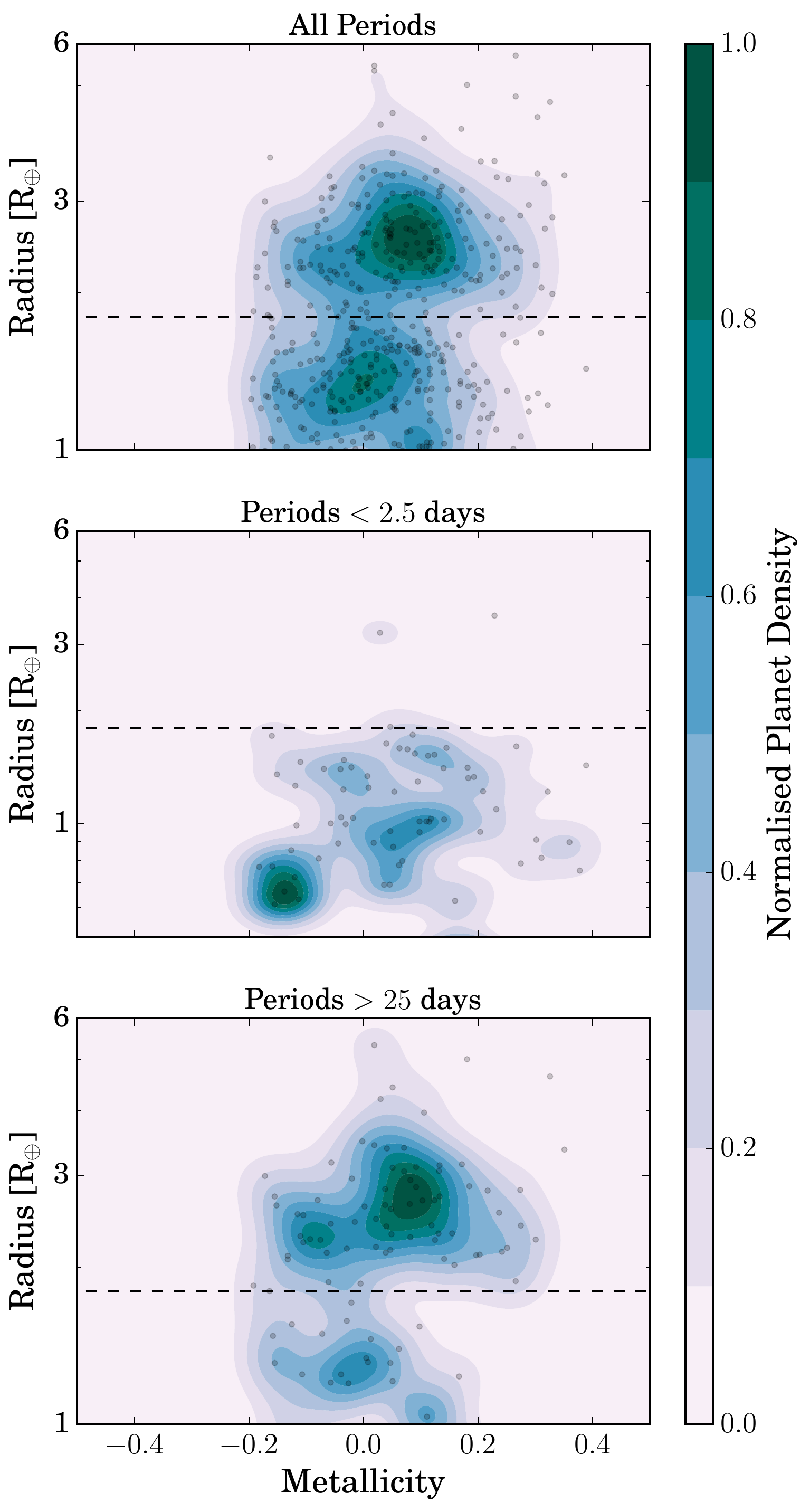}
\caption{Same a Figure~\ref{fig:period_cut_panels}, but only for planets whose host stars have masses in the range 0.85-1.15~M$_\odot$ {\bc and metallicities $>-0.2$}.}\label{fig:mass_cut1}
\end{figure}
The result is shown in Figure~\ref{fig:mass_cut1}. It is clear that the trend that ``large" planets are bigger around more metal-rich stars still holds true in this sub-sample, both at all periods and long periods. Small number statistics prevent us confirming the trend that at short periods, small planets are larger around more metal rich stars. In order to investigate this further, we alternatively cut the sample into a narrow metallicity range with a spread of $\pm 0.1$dex around the median value and compare the radius distribution for planets around high and low mass stars (split at the median stellar mass value of 1.02~M$_\odot$). 
Small number statistics prevent us from making robust statements, but there is no statistically significant evidence of a clear trend with stellar mass. This gives us cautious optimism to presume that the significant trend of bigger (more massive) solid cores around higher metallicity stars is robust. 

We need to be more cautious about the trend that small planets at long periods are more frequent around lower metallicity stars. 
There is weak, but not statistically robust evidence, that both sub-samples covering a narrow range in metallicity, but wider range in mass or a narrow range in stellar mass, but a wider range in metallicity show the trends that could be responsible for driving the original observations that small planets are more frequent at long periods around CKS stars with lower metallicities. However, here we do not have the sample size to be confident about the driver; one would suspect that both are playing a part. 

Additionally, one might worry that because high-metallicity stars are more massive (and hence brighter) in the CKS sample, small planets at large periods are harder to observe around these objects.  \citet{Petigura2018}, in measuring absolute occurrence rates, confirm that at periods of 10-100 days, small planets are more abundant around low-Z stars in both an absolute and relative sense (while larger planets are more common around high-Z stars), indicating that the trend we identify is not coming from this selection bias.

\begin{figure}
\centering
\includegraphics[width=\columnwidth]{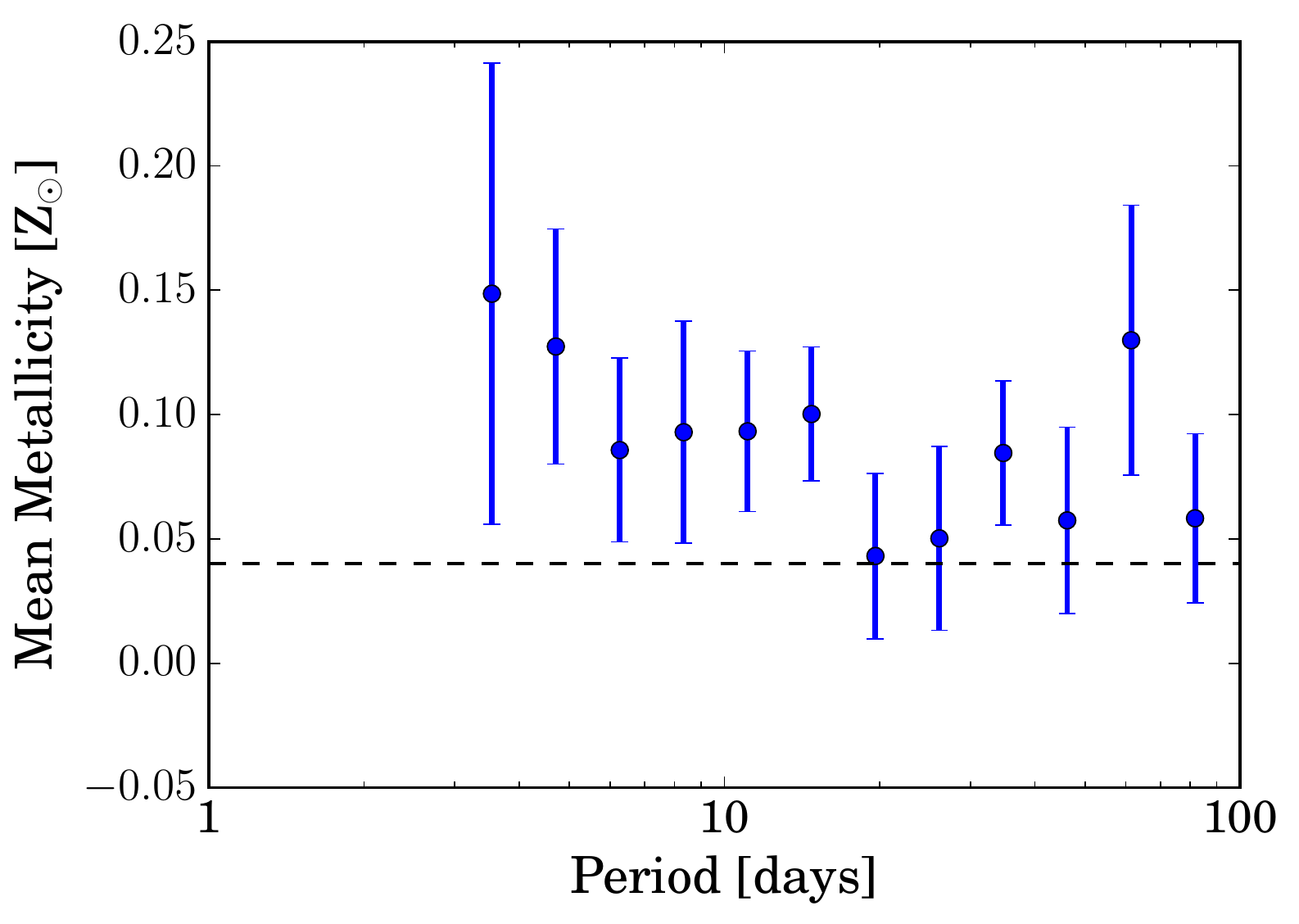}
\caption{Same as Figure~\ref{fig:Av_met_period}, except this time just for those planets whose host stars have masses in the range 0.85 to 1.15 M$_\odot$ {\bc and metallicities $>-0.2$}.}\label{fig:Av_met_cut}
\end{figure}

Finally, we reproduce Figure~\ref{fig:Av_met_period} in Figure~\ref{fig:Av_met_cut} for our narrow stellar mass range and see indeed that {\bc the} trend identified previously is indeed robust to the correlation between stellar mass and metallicity.  

In summary, we have checked to see whether the trends identified previously in Sections 3 and 4 are robust to the correlation between stellar mass and metallicity present in the \citet{Fulton2017} sample of CKS stars. We find no evidence that metallicity is just a proxy for stellar mass in driving these trends; except in the case where long-period, smaller planets are more frequent around metal poor CKS stars. This may be driven by metallicity, stellar mass, or a combination of the two.




\section{Implications for Planet Formation}\label{sec:discuss}

In the following section we discuss the implications and interpretation of our results in the context of planet formation. We will be careful to make sure those implications that do or do not depend on the interpretation and origin of the radius gap are made clear. 

The ``evaporation-valley'' interpretation of the gap in the radius distribution is the best established theory as such a gap was previously predicted \citep{OwenWu13,Lopez2013}. A key indication of an evaporative origin of the radius-period valley would be that the planetary radius of the gap declines with period, as more massive cores can only be stripped at short periods. \citet{Fulton2017}, were unable to confirm the negative slope of the valley as the planetary radii were not precise enough. A declining gap radius with period was recently confirmed using a sample of planets for which asteroseismic determined stellar parameters allowed for a precise identification and measurement of the radius-period valley. \citet{vaneylen2017} showed that the valley radius varied with period as $R_{\rm valley}\propto P^{m}$ with $m=-0.09^{+0.02}_{-0.04}$. A further prediction of the evaporation valley theory is that large planets should be rarer at high bolometric isolations around lower mass stars \citep{OwenWu13}. Therefore, as the evaporative origin of the radius gap has yet to be fully tested against all its predictions, in this work we will be conservative and shall not assume that it is the only explanation \citep[e.g.][]{Ginzburg2018}. 

Indeed, the same physics that explains the origin of the evaporative-valley can be made to work in reverse. Since it only takes a small addition of mass to have a large enough envelope to appear as a large planet and ``hop" accross the gap; such an addition of mass is easy to accrete. In the context of planetary cores embedded in a gaseous disc this means it only requires a small change in core mass to change from a core that accretes a small atmosphere, thus appearing as a small planet, from one that accretes a slightly larger atmosphere and appears as a large planet. The core radius (mass) and period dependence of such a hypothesis have not been worked out in any detail and are likely to be sensitive to the local disc conditions. For example, the ``standard'' assumptions made for the models of \citet{Lee2015}, would yield a gap that is inconsistent with the CKS results, with it appearing around {\bc a core mass of $\sim$ 1~M$_\oplus$ (their Figure~4), placing it at a radius of approximately 1~R$_\oplus$}. However, an increase in the gas metallicity by a factor of $\sim 100$ {\bc above solar} (e.g. through an enhancement of the dust-to-gas ratio) can increase the radius of the gap to match the observations {\bc within the \citet{Lee2015} framework}. Given the uncertainty in the dust-to-gas ratio, especially in the inner regions of protoplanetary discs, such increases are perfectly plausible, as radial-drift can preferentially enhance the dust-to-gas ratio in the inner regions of discs \citep{Birnstiel2012}.   

Our results for the small planets at short periods and large planets at longer periods are essentially in agreement. {\bc We assume} the small planets at short periods probe the properties of the solid cores and the large planets at large periods probe the primordial H/He envelopes the solid cores accreted while in the gas disc. Both these samples of planets are larger around higher metallicity stars. The amount of nebula gas a core can accrete is essentially set by {\bc three} things: the depth of the core's gravitational potential well, how quickly the accreting gas can cool {\bc and the mean-molecular weight of the gas}. Higher metallicity gas has a higher opacity and therefore accreting envelopes cannot cool as efficiently: a fixed core mass surrounded by higher metallicity gas should result in a smaller bound envelope. For the case of H/He dominated atmospheres in the inner disc, where the cooling is set by radiative-convective boundary occurring when Hydrogen dissociates, \citet{Lee2015} show that the envelope-mass fraction scales with envelope metallicity as $\propto Z_{\rm env}^{-0.4}$. {\bc (For $Z\gtrsim0.2$ this slower cooling can be counteracted by the decrease in the envelope's scale height for  envelopes heavily enriched in metals due to the gas's higher mean-molecular weight.)} Unsurprisingly, the amount of gas a planet can accrete is more sensitive to the depth of the potential well, with more massive cores accreting larger envelopes. Since our results indicate that higher metallicity stars host planets with more massive atmospheres then we would expect them to have more massive cores {\bc or enriched envelopes} as well, allowing them to overcome the slower cooling of higher metallicity gas. We do find that core mass probably increases with stellar metallicity, when we demonstrated the solid cores at short periods were indeed larger, and presumably more massive, around more metal rich stars. Therefore, we conclude the higher solid content in protoplanetary discs around higher metallicity stars results in more massive solid cores which subsequently accrete larger gas envelopes. 

This result is intuitively reasonable---discs with more solid material produce larger solid planets---but we note that theoretically speaking, it is not obvious that this would be the case.  Higher metallicity stars are known to host more gas giants at short orbital periods \citep{FischerValenti} and are likely more prone to produce gas giants at larger periods as well.  Gaps opened by these planets in the gas disc produce pressure maxima, which can trap pebble-sized solids as they drift toward the host star \citep{Rice2006}.  This dust filtration mechanism has been proposed as an explanation for observed ``transition discs" \citep[reviewed in][]{EspaillatPPVI}, which host inner cavities cleared of dust. These transition discs have large dust holes ($>10$ AU) and high accretion rates and may host giant planets \citep[see][]{Owen2016} at wide separations.  This idea has some observational motivation as the photospheres of Herbig Ae/Be stars that host transition discs appear to be depleted in refractory elements \citep{Kama2015}.  The pressure traps in the disc created by the giant planets are then preventing the inward drift of dust particles, so refractory-depleted gas is accreting onto the surface of the Herbig stars.  This result suggests that the presence of a giant planet could starve the region close to the star of the solid material needed to produce large cores.  In other words, if higher metallicity stars are more likely to produce giant planets at larger periods, which trap solids, preventing them from accumulating in the inner disc, high metallicity stars could be less likely to host close-in planets.  Our results are not consistent with this story.

In addition to the finding that cores are smaller around low metallicity stars at both small and larger periods, we have shown that small planets are more frequent (in a relative sense, see \citealt{Petigura2018} for the absolute calculations) around lower metallicity (or lower mass) stars at periods {\bc $>$25 days.}  We interpret this to mean that when more solids are present, it is easier to {\bc produce cores that can} hold onto a H/He envelope {\bc before the gas disc dispersed}\footnote{This should not be confused with the critical core mass required to undergo runaway accretion and become a gas giant which requires already {\bc accreting an envelope mass comparable to the core mass, such that self-gravity of the envelope causes it to contract as more mass it added.}}. Therefore, these small planets at long periods are likely to be those where the solids could not produce a core massive enough to acquire an envelope with a mass fraction $\gtrsim 0.01$ before the gas disc dispersed \citep{OwenWu17}. {\bc Alternatively, if stellar XUV luminosity were a strong function of stellar metallicity, especially at early times, then this could play a part in driving this trend. For example, if lower metallicity stars were more XUV luminous.}  

All of the above are independent of the assumptions as to the origin of the radius gap. However, it is useful to go further and make use of our theoretical understanding as to the origin of the radius-period valley to highlight additional important, interesting,  but speculative conclusions that we can draw about the planet formation process. 

The photoevaporative origin for the radius gap {\bc{\it assumes}} that planets grow massive enough to accrete gas envelopes before the gas disc disperses and that the bulk of the small planets are those planets for which evaporation has removed their voluminous primordial H/He atmospheres \citep{OwenWu17}. In this {\bc simplistic scenario} no small planets at long periods would be expected, as photoevaporation is ineffective at completely removing an atmosphere at large periods. Now clearly the formation history of the Solar-System indicates that not {\it every} solid planet formed through the above pathway and that a second population of ``born-terrestrial'' planets can form on a longer timescale after the gas disc disperses. \citet{OwenWu17} discuss tentative evidence for this second population as a minor constituent. 

\begin{figure}
\centering
\includegraphics[width=\columnwidth]{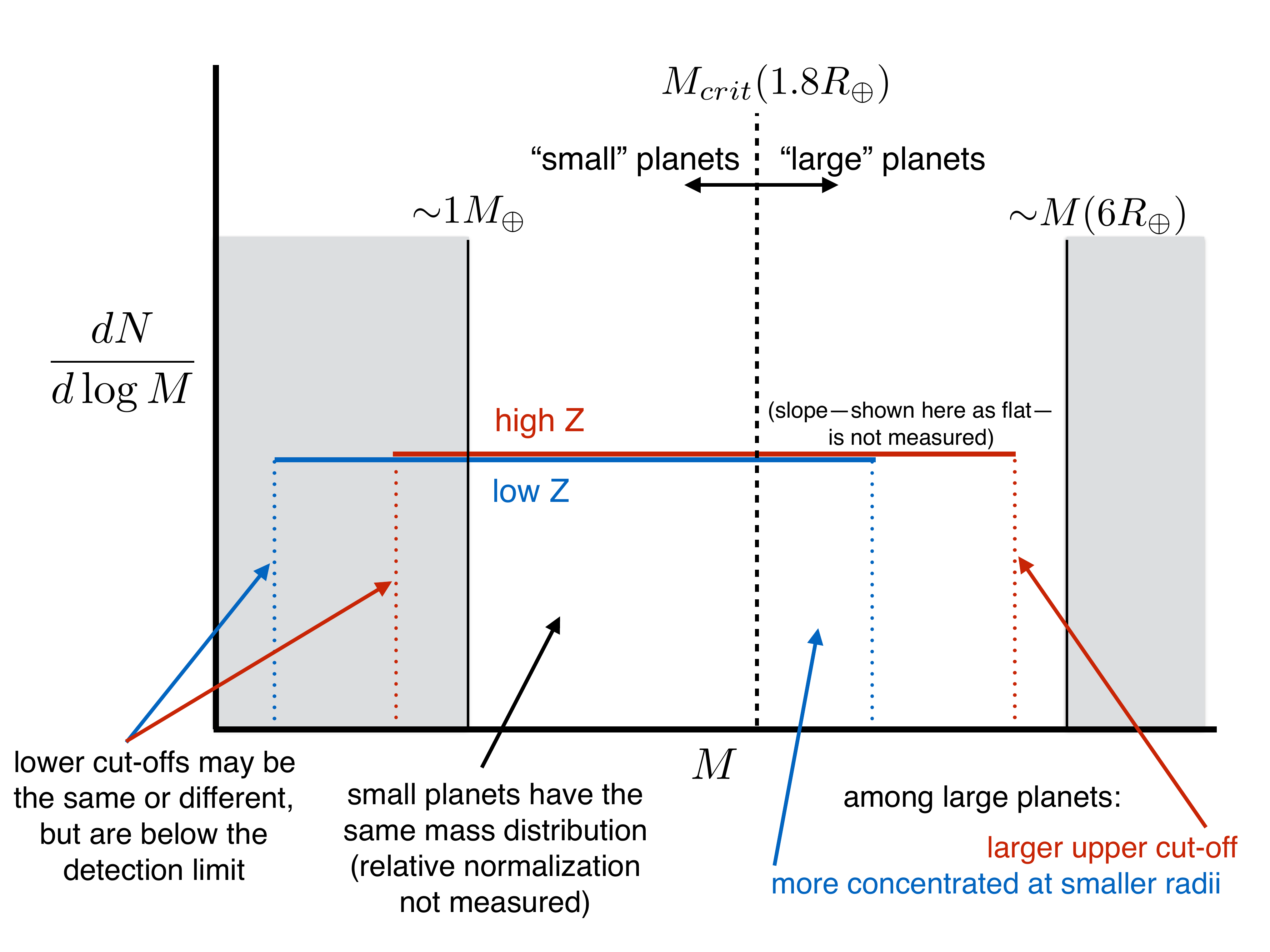}
\caption{Schematic diagram demonstrating that several salient features of the observations could be matched if both high- and low-metallicity stars host planets with the same functional form for their mass distribution but the upper mass limit is higher for high-metallicity stars. We are assuming, as described above, that the transition radius across the gap occurs when the core mass is sufficiently massive to have accreted a $\sim 1$\% H/He envelope. The parameter $dN/d\log M$ is the number of planets produced per log bin in mass.  We have not constrained the slope of this function and display it as flat (equal numbers of planets at each mass scale) for simplicity. The relative normalization of the mass distribution between high-Z and low-Z stars is likewise unconstrained in this work because we do not have information about the metallicity distribution of non-planet-hosting stars. The exoplanet properties at large periods show two features that are consistent with this schematic model: the radii of ``large" planets extend to larger values around high-metallicity stars, and the radius distribution for ``small'' planets is the same around low- and high-metallicity stars.}\label{fig:size_distrib_diagram}
\end{figure}

Before appealing to two formation epochs (or, more likely, planet formation occurring over an extended time period), it is worth evaluating whether a reasonable size distribution of planet masses produced at a similar epoch could reproduce the size distribution of planets at periods between 25 and 100 days. 
As illustrated in Figure~\ref{fig:size_distrib_diagram}, if planets around low and high metallicity stars have occurrence frequencies that scale with mass in the same way but have different upper mass cut-offs, high metallicity stars will be observed to host, on average, ``large" planets with larger radii but ``small'' planets with the same radius distribution than their low metallcity counterparts (so long as the lower end of the planetary mass distribution remains unobserved).  This interpretation has the nice feature that it explains the otherwise-puzzling metallicity-independence of small planet radii at large periods.

In this scenario, should there be a distinction between small, bare planets and large, gas-rich planets?  In general, yes: the fact that only a small amount of atmospheric gas is necessary to dramatically increase a planet's radius, invoked to explain the evaporation valley, could also be used to argue for a planetary radius gap produced as a result of formation.  Such a ``formation valley" would result from the point where the cores become massive enough to accrete 1\% atmospheres and appear as large planets. In this case close-in planets form as a single population in the gas disc and only those that are massive enough accrete a gas envelope. This transition mass is somewhere in the range of $\sim3-5$M$_\oplus$ corresponding to the mass of Earth-like cores where the radius distribution of small planets starts to decline. 

However, there are two problems with interpreting the 1.8\,R$_\oplus$ radius gap as a formation valley.  Firstly, the mass at which planetary cores ``hop'' across the gap is not observed to be a strong function of stellar metallicity. Since the accretion of atmospheres depends on how fast such an atmosphere can cool {\bc and the mean-molecular weight of the gas} this is evidently metallicity dependent, implying the hopping mass must be metallicity dependent.  However, how weak or strong this dependence is in reality remains open. {\bc For example,} using the models of \citet{Lee2015} we find a core-mass scaling $\propto Z_{\rm env}^{0.22}$, for {\bc a core-mass--radius scaling of $M_c\propto R_c^4$}; but of course it is unclear how to map the stellar metallicity onto a forming planet's atmospheric metallicity. Second---and perhaps more importantly---the gap is observed throughout the full period range extending to 100 days \citep{Fulton2017}, and the radius corresponding to the upper edge of the gap does not vary substantially with period (see Figure \ref{fig:rad_period_met_bins}).

{\it Now within the evaporation-valley hypothesis}, the fact that small planets at long periods are more prevalent (in a relative sense) around lower metallicity stars requires a population formed after the disc disperses. Since the size of these planets (and therefore assumed mass) is high enough to have accreted a significant gas envelope if the gas disc remained \citep[e.g.][]{Rafikov2006}, they should have formed after the gas disc dispersed.  This late-stage formation process makes use of smaller constituents which were large enough to survive  gas disc dispersal (e.g. planetesimal sized or bigger). Since these planets are prevalent around lower metallicity stars, which presumably had a lower solid content, {\bc we hypothesise} it was harder to turn this solid material from planetesimal/embryo-sized objects into cores massive enough to accrete gas. This scenario is consistent with the interpretation that total solid content in the disc determines  the production rate and ultimate masses of solid cores before the gas disc disperses, with higher solid contents resulting in more massive cores \citep[e.g.][]{Kokubo2006} as indicated by the larger cores found around higher metallicity stars. Since, in the pure evaporation-valley scenario, the properties of the observed radius gap imply that the most common core mass is around 3~M$_\oplus$, a significant mass range exists between cores that cannot accrete an envelope and the $\sim$3~M$_\oplus$ peak in the distribution of cores that did grow large enough to accrete an envelope.  In this interpretation, the final bottleneck in the growth of solids thus sits somewhere between planetesimal sizes and sub-Earth sized solid bodies.  If planets reach Earth mass during the lifetime of the gas disc, they grow into {\it Kepler}-like planets.  Formation of 1-3 Earth-mass cores is limited to late-stage, gas-free growth, such as the giant impact phase long posited for the formation of the Earth.  

We conclude that the data can also be plausibly matched by a scenario with no formation valley in the following way: if planet formation in the gas disc experiences a bottleneck between planetesimal and Earth-sized bodies and results in a core mass distribution peaked around 3~M$_\oplus$, then photoevaporation can reproduce the lower-radius cutoff for this population.  Most 1-3~R$_\oplus$ planets that do not host substantial gas envelopes then form during a late-stage giant impact phase.  We note that in this scenario, while the gap radius should be metallicity-independent (as observed), the width of the gap should depend on stellar metallicity.  This is because the maximum mass solid planet produced during the giant impact phase should be larger for discs containing more solids \citep[e.g.][]{Kokubo2006}.  Though this is inconsistent with our reported lack of metallicity-dependence for the sizes of small distant planets, a subtle metallicity dependence cannot be ruled out by current data.

We believe that the formation valley and late stage formation scenarios discussed above for explaining the radius distributions at large periods will be distinguishable using data from Plato which will yield larger numbers of longer period planets.  {\bc If this formation hypothesis of the radius gap is correct}, the period dependence of the gap radius should exhibit a break at periods beyond $\sim$25 days.  If late stage terrestrial planet formation dominates, no break should appear in the period dependence of the upper limit of the radius gap, but better data should reveal larger long-period gas-free planets around higher metallicity stars.  We have emphasized stellar metallicity dependence, which can be probed effectively in the CKS sample, but we note that in future observations, host star mass may provide a better lever arm for distinguishing between these scenarios in the near future. 

In summary, it is clear that the stellar metallicity, which we take as a proxy for the total solid content in the original protoplanetary disc, imprints itself in the population of close-in, low-mass planets. The precision of the California-{\it Kepler}-Survey stellar properties has finally allowed us to reveal important trends. Specifically, we find higher solid content discs can clearly grow more massive cores, which can consequently accrete more massive atmospheres, even in the face of the slower cooling of these accreting atmospheres. All these trends are consistent with the general picture of core accretion. Additional exciting implications about how discs turn solids into larger bodies are inextricably linked to the physical origin of the occurrence valley in the radius-period distribution. The ``evaporation-valley'' model is clearly favoured at short periods.  At long periods, a ``formation valley" may also be present.  If not, our results {\bc suggest} that discs can readily form solid bodies planetesimal sized or larger, but it requires a significant solid reservoir to turn these into solid cores. Testing key predictions of the evaporation scenario, such as the more efficient evaporation around later-type stars at fixed bolometric isolation \citep{OwenWu13} is imperative.

\section{Summary and Conclusions}\label{sec-sum}

In this work we have used the results from the California-{\it Kepler}-Survey to study how close-in, low-mass planets vary with their host star's metallicity. By making sure to account for how photoevaporation will sculpt the exoplanet population over its lifetime we have been able to extract the imprints of host star metallicity on planetary properties and the associated implications on the origin of the ubiquitous close-in exoplanets. Our main results are as follows:
\begin{itemize}
\item {\bc We find planets at short periods are larger around higher metallicity stars. At short periods photoevaporation will have enitrely stripped any H/He envelope from the core. Therefore we interpret this results as the evidence that solid cores of close-in planets are larger and therefore more massive around higher metallicity stars. }
\item {\bc Planets at long periods which are thought to host H/He atmospheres are larger around higher metallicity stars. At long periods photoevaporation is ineffective at removing significant mass. Thus we expect planets around higher metallicity stars accrete larger H/He envelopes. As higher metallicity H/He envelopes are harder to accrete due to slower cooling, this must be counteracted by either larger core masses {\bc or heavily enriched atmospheres} at higher metallicity.}
\item H/He atmosphere hosting planets are more common around higher metallicity stars at short periods, where the average host star metallicity begins to increase at a period of $\sim 20$~days. This is interpreted as what is expected from photoevaporation, as higher metallicity exospheres will be cooler due to enhanced cooling from atomic metal lines and therefore will drive less efficient photoevaporative outflows. 
\item {\bc We find} small ($<1.8$~R$_\oplus$) terrestrial planets are more common around low metallicity stars at long periods. If the bulk of terrestrial planets are photoevaportively stripped cores that accreted voluminous H/He atmospheres at birth then this population must have formed after the gas disc dispersed. Therefore, terrestrial-like planet formation could be more prevalent around lower-metallicity stars. This inference indicates that discs with a lower metal content struggled to form cores with masses $\gtrsim 1$~M$_\oplus$ before the gas disc dispersed, yet were able to grow to such masses after dispersal, from planetesimals/planetary embryos that remained.
\end{itemize}


\section*{Acknowledgements}
We are grateful to the referee for a helpful report that improved this paper. We thank BJ Fulton, Erik Petigura and Yanqin Wu for comments on earlier drafts of the paper. JEO is supported by a Royal Society University Research Fellowship and grateful to UCSC for hospitalility during the completion of this work. RMC is supported by NSF CAREER grant  AST-1555385. We are grateful to the CKS team for useful discussions and making their data publicly available.

\end{document}